\documentclass[12pt]{iopart}

\usepackage{epsfig, color, psfrag, graphics}
\usepackage{amsthm, bm, amsfonts, ae}

\newcommand{\upd}{\mathrm{d}}
\newcommand{\bvec}[1]{{\bf\string#1 }}

\begin{document}

\title[Density functional theory for hard-sphere mixtures]{Density functional
  theory for hard-sphere mixtures: the White-Bear version Mark II}

\author{Hendrik Hansen-Goos\dag\ddag and Roland Roth \dag\ddag}

\address{\dag\ Max-Planck-Institut f\"ur Metallforschung - Heisenbergstr. 3,
 70569 Stuttgart, Germany}
\address{\ddag\ ITAP, Universit\"at Stuttgart - Pfaffenwaldring 57, 70569
 Stuttgart, Germany}

\ead{hhansen@fluids.mpi-stuttgart.mpg.de}

\date{\today}

\begin{abstract}
In the spirit of the White-Bear version of fundamental measure theory we
derive a new density functional for hard-sphere mixtures which is based on a
recent mixture extension of the Carnahan-Starling equation of state. In
addition to the capability to predict inhomogeneous density distributions very
accurately, like the original White-Bear version, the new functional improves
upon consistency with an exact scaled-particle theory relation
in the case of the pure fluid. We examine consistency in detail within the
context of morphological thermodynamics. Interestingly, for the pure fluid the
degree of consistency of the new version is not only higher than for the
original White-Bear version but also higher than for Rosenfeld's original
fundamental measure theory. 
\end{abstract}

%\pacs{}

\section{Introduction}

In the mid 1970s, density functional theory, which was originally formulated
for quantum systems, has been extended to systems that follow classical
statistical mechanics \cite{Ev79}. Since then, density functional theory of
classical systems (DFT) has developed to an indispensable tool for the study
of  inhomogeneous  systems such as crystals, fluids in confined geometries
\cite{Ev90}, liquid-vapor interfaces and wetting and drying on substrates (for
a recent review see \cite{Wu06}). DFT is based on the fact that there exists a
functional $\Omega[\rho]$ of the spatially varying particle number density
$\rho(\bvec{r})$ which possesses two properties: (i) it is minimized by the
equilibrium density $\rho_0(\bvec{r})$, and (ii) the minimum value
$\Omega[\rho_0]$ equals the grand potential $\Omega$ of the system. These
properties give rise to the variational principle $\delta \Omega[\rho]/\delta
\rho \equiv 0$ for $\rho(\bvec{r}) = \rho_0(\bvec{r})$. 

One can decompose $\Omega[\rho]$ as 
\begin{equation}
\label{eq_gpfunc}
  \Omega[\rho] = \mathcal{F}[\rho] + \int\! \upd \bvec{r} \,\rho({\bf r})
  (V_{\mathrm{ext}}(\bvec{r}) - \mu) \, ,
\end{equation}
where $\mu$ is the chemical potential, $V_{\mathrm{ext}}(\bvec{r})$ the
external potential acting on the particles and $\mathcal{F}[\rho]$ is a
unique functional corresponding to the intrinsic Helmholtz free energy of the
system in equilibrium.

In principle, Eq.~(\ref{eq_gpfunc}) with the variational principle constitutes
an excellent tool for (numerical) calculations of $\rho_0(\bvec{r})$ and hence
the grand potential $\Omega[\rho_0(\bvec{r})]$ in arbitrary external
potentials. Unfortunately, for many systems of interest, only more or less
crude approximations for $\mathcal{F}[\rho]$ are known. The expression for the
ideal gas, however, is known exactly
\begin{equation}
  \beta \mathcal{F}_{\mathrm{id}}[\rho] = \int \! \upd \bvec{r} \, \rho(\bvec{r}) \left(\ln
  (\Lambda^3\rho(\bvec{r})) - 1\right) \, ,
\end{equation}
where $\Lambda$ is the thermal wavelength of the particles and $\beta =
1/(k_{\mathrm{B}} T)$ with Boltzmann's constant $k_{\mathrm{B}}$ and the
temperature $T$. The interactions among particles are described by the excess
(over ideal gas) free energy $\mathcal{F}_{\mathrm{ex}}[\rho] =
\mathcal{F}[\rho]-\mathcal{F}_{\mathrm{id}}[\rho]$ which for our purposes can
be expressed as $\beta\mathcal{F}_{\mathrm{ex}}=\int\!\upd \bvec{r} \Phi(\bvec{r})$, where the excess free energy density $\Phi(\bvec{r})$ is a
functional of $\rho(\bvec{r})$.

In this work we direct our attention to mixtures of hard-spheres. For these
systems, the exact expression for $\mathcal{F}_{\mathrm{ex}}$ is unknown but a
number of approximations can be found in the literature \cite{Ev92}. The
interest in the hard-sphere system is manifold. The hard-sphere system serves
as a reference system for fluids with short-ranged repulsion and additional
attractive interactions among
particles. The attractive part of the potential is usually treated
perturbatively \cite{HaMcDo}. Furthermore, colloidal suspensions with quasi
hard-sphere interactions can be realized experimentally (see, e.g.\
\cite{RuEA96}) which provides a test ground for predictions from the field of
the purely entropically driven hard-sphere systems, like entropic forces
\cite{RoEvDi00}, asymptotic decay of correlation functions in mixtures
\cite{GrEA04,GrEA05} etc..

A very successful class of excess free energy functionals is formulated
within the framework of fundamental measure theory (FMT) introduced by
Rosenfeld \cite{Rf89}. While the original FMT has the Percus-Yevick (PY)
equation of state as an output, later, with the setup of the White-Bear
version of FMT \cite{RoEvLaKa02,YuWu02}, the more accurate Carnahan-Starling
(CS) equation of state was incorporated into FMT. The resulting gain in
precision in the structure of inhomogeneous density distribution
\cite{RoEvLaKa02,YuWu02,BrRoScDi03} and in thermodynamics, however, has to be
paid for with a slight inconsistency appearing on the level of the pressure
\cite{RoEvLaKa02}: the pressure in the hard-sphere fluid obtained from a
scaled-particle theory equation differs slightly from the underlying bulk equation of
state. The aim of this work is to build upon the White-Bear version of FMT,
using a new mixture formulation of the CS equation of state \cite{HaRo06},
such that this inconsistency is resolved.

The paper is organized as follows. In Sec.~\ref{sec_fmt} we review Rosenfeld's
derivation of FMT. Section~\ref{sec_wbII} is dedicated to the presentation of
the White-Bear version of FMT and the derivation of the new version of FMT. In
Sec.~\ref{sec_mtd} we give a brief introduction to morphological
thermodynamics of hard-sphere fluids and compare the performance of
Rosenfeld's FMT and the original White-Bear version with that of the new
functional. Section~\ref{sec_concl} contains our conclusion. 

\section{Fundamental measure theory}
\label{sec_fmt}

In a seminal paper in 1989, Rosenfeld set up FMT, allowing him to derive his
successful free energy functional for the hard-sphere mixture \cite{Rf89}. We
sketch his approach briefly in the following.

We consider a $\nu$-component hard-sphere mixture with spatially varying
particle number densities $\rho_i(\bvec{r})$, $i=1,\ldots,\nu$. From the
theory of diagrammatic expansions \cite{HaMcDo} the excess free energy
functional in the low-density limit is known exactly. One finds that the
Mayer-$f$ function, related to the pair interaction potential $V_{ij}(r)$
between the particles of species $i$ and $j$ by $f_{ij}(r) =
\exp[-\beta V_{ij}(r)]-1$, plays a central role. In the case of hard spheres,
$V_{ij}$ is either infinite if spheres overlap and zero otherwise. As a result
the Mayer-$f$ function obtains a purely geometrical meaning:
$f_{ij}(r) = -\Theta(R_i+R_j-r)$ where $R_i$ and $R_j$ are the radii of the
respective species and $\Theta$ is the Heaviside function.

Inspired by the exact excess free energy functional of the one-dimensional
hard-rod mixture \cite{VaDaPe89,Rf90}, the key-idea of FMT is the
deconvolution of the Mayer-$f$ function $f_{ij}(r)$ into a sum of products
with factors depending only on one of $R_i$ and $R_j$. Rosenfeld's
deconvolution reads 
\begin{equation}
\label{eq_deconv}
\hspace{-2.0cm}
  -f_{ij}(|\bvec{r}_i-\bvec{r}_j|) = \omega_0^i \otimes \omega_3^j + 
  \omega_3^i \otimes \omega_0^j + \omega_1^i \otimes \omega_2^j + \omega_2^i
  \otimes \omega_1^j - \boldsymbol{\omega}_1^i \otimes \boldsymbol{\omega}_2^j
  - \boldsymbol{\omega}_2^i \otimes \boldsymbol{\omega}_1^j 
\end{equation}
with four scalar and two vectorial (weight) functions
\begin{equation}
  \hspace{-1.0cm} \omega_3^i(\bvec{r}) =  \Theta (R_i-|\bvec{r}|) \, , \quad
  \omega_2^i(\bvec{r}) =  \delta (R_i-|\bvec{r}|) \, , \quad
  \boldsymbol{\omega}_2^i(\bvec{r}) =  \frac{\bvec{r}}{|\bvec{r}|}
  \,\delta(R_i-|\bvec{r}|) \, ,
\end{equation}
and $\omega_1^i(\bvec{r})=\omega_2^i(\bvec{r})/(4 \pi R_i)$,
$\omega_0^i(\bvec{r})=\omega_2^i(\bvec{r})/(4 \pi R_i^2)$, and
$\boldsymbol{\omega}_1^i(\bvec{r})=\boldsymbol{\omega}_2^i(\bvec{r})/(4 \pi
R_i)$. The convolution product $\otimes$ in Eq.~(\ref{eq_deconv}) is defined by
\begin{equation}
  \omega_{\alpha}^i \otimes \omega_{\gamma}^j = \int\!\upd\bvec{r}\,
  \omega_{\alpha}^i(\bvec{r}-\bvec{r}_i)\cdot\omega_{\gamma}^j(\bvec{r}-\bvec{r}_j)
  \, ,
\end{equation}
where the dot $\cdot$ stands for the usual product in the case of the scalar
weight functions and for the scalar product in the case of the vectorial weight
functions. Using the weight functions, one can define weighted densities
\begin{equation}
\label{eq_weighden}
  n_{\alpha}(\bvec{r}) = \sum_{i=1}^{\nu} \int\!\upd \bvec{r}'
  \rho_i(\bvec{r}') \omega_{\alpha}^i(\bvec{r}-\bvec{r}') \, .
\end{equation}

The deconvolution, Eq.~(\ref{eq_deconv}), can be used to express the exact 
low-density limit of the excess free energy functional:
\begin{equation}
\label{eq_ldlimitwd}
\eqalign{
  \lim_{\rho_i\to 0}\beta \mathcal{F}^{\mathrm{ex}} & = - \frac{1}{2}
  \sum_{i,j=1}^{\nu} \int\!\upd\bvec{r}\upd\bvec{r}' \rho_i(\bvec{r})
  \rho_j(\bvec{r}') f_{ij}(|\bvec{r}-\bvec{r}'|) \\
  & = \int\!\upd\bvec{r} \big(
  n_0(\bvec{r}) n_3(\bvec{r}) + n_1(\bvec{r}) n_2(\bvec{r}) -
  \bvec{n}_1(\bvec{r}) \cdot
  \bvec{n}_2(\bvec{r}) \big) \, .
}
\end{equation}

This result, together with the structure of the exact excess free energy
functional of one-dimensional hard-rod mixtures, leads to the assumption that
the excess free energy density $\Phi(\mathrm{r})$ can be approximated as a
function of the six weighted densities only. This assumption guarantees to
recover the exact low density limit of $\Phi(\mathrm{r})$. The expression for
the free energy density is obtained by an extrapolation of the known
low-density result for $\Phi(\mathrm{r})$ to higher densities using
thermodynamic arguments.

We consider the case of a homogeneous hard-sphere mixture, i.e.\ the density
distributions $\rho_i(\bvec{r})\equiv \rho_i=N_i/V$ are constant. $N_i$ is the
number or spheres of species $i$ in the volume $V$. The excess pressure
$p^{\mathrm{ex}}$ of a fluid mixture can be obtained from the excess free
energy density $\Phi$ via
\begin{equation}
  \beta p^{\mathrm{ex}} = - \frac{\partial(V\Phi)}{\partial V} = -\Phi +
  \sum_{i=1}^{\nu} \frac{\partial \Phi}{\partial \rho_i} \rho_i = -\Phi +
  \sum_{\alpha} \frac{\partial \Phi}{\partial n_{\alpha}} n_{\alpha} \, .
\end{equation}
Note that the vectorial weighted densities, which are formally included in the
above sum, actually vanish in the uniform fluid. The ideal gas contribution to
the pressure is $\beta p^{\mathrm{id}} = \sum_i \rho_i$, which in terms of the
weighted densities reduces to $\beta p^{\mathrm{id}} = n_0$. Hence, according
to thermodynamics (TD), the total
pressure within FMT can be written as 
\begin{equation}
\label{eq_pTD}
  \beta p_{\mathrm{TD}} = n_0 -\Phi + \sum_{\alpha} \frac{\partial
  \Phi}{\partial n_{\alpha}} n_{\alpha} \, .
\end{equation}

On the other hand, there is an exact relation from scaled-particle (SP) theory (see
\cite{Rf89} and references therein) between the chemical potential $\mu_i$ of
species $i$ for a very large sphere and the reversible work required for the
creation of a cavity that can hold the large sphere of species $i$: in the
limit $R_i\to \infty$ one obtains $\mu_i/V_i \to p_{\mathrm{SP}}$, where
$p_{\mathrm{SP}}$ is the total pressure of the fluid mixture and $V_i=
(4/3)\pi R_i^3$. In our context, this is equivalent to
(cf. \cite{Rf89,RoEvLaKa02})  
\begin{equation}
\label{eq_pSP}
  \beta p_{\mathrm{SP}} = \frac{\partial \Phi}{\partial n_3} \, .
\end{equation}

Obviously, we obtain a differential equation for $\Phi$ by equating the
expressions for $p_{\mathrm{TD}}$ and $p_{\mathrm{SP}}$. This equation was
solved by Rosenfeld who used the ansatz
\begin{equation}
\label{eq_rfansatz}
\hspace{-1.5cm}
\Phi = f_1(n_3) n_0 + f_2(n_3) n_1 n_2 + f_3(n_3) \bvec{n}_1\cdot\bvec{n}_2 +
f_4(n_3) n_2^3 + f_5(n_3) n_2 \bvec{n}_2\cdot\bvec{n}_2 \, ,
\end{equation}
with $f_1, \ldots, f_5$ being functions of the dimensionless weighted density
$n_3$. The ansatz Eq.~(\ref{eq_rfansatz}) combines all multiplicative
combinations of the weighted densities which share the dimension of $\Phi$,
i.e.\ $\mathrm{(length)}^{-3}$. There is a solution $\Phi_{\mathrm{RF}}$ of the
SPT differential equation. The integration constants can be fixed by the
following additional requirements: (i) in the low-density limit,
Eq.~(\ref{eq_ldlimitwd}) is recovered, (ii) for the one-component uniform 
hard-sphere fluid the correct third virial coefficient is reproduced and (iii)
the pair direct correlation function $c^{(2)}(r)$ is regular for $r \to 0$,
which enforces the prefactor for the term $\propto n_2
\bvec{n}_2\cdot\bvec{n}_2$. The result is 
\begin{equation}
\label{phi_RF}
  \Phi_{\mathrm{RF}} = - n_0 \ln (1-n_3) + \frac{n_1 n_2 - \bvec{n}_1 \cdot
   \bvec{n}_2}{1-n_3} + \frac{n_2^3 - 3 n_2 \bvec{n}_2 \cdot
   \bvec{n}_2}{24\pi(1-n_3)^2} \, .
\end{equation}

Rosenfeld's excess free energy gives a good account for many aspects of
nonuniform hard-sphere fluids, pure \cite{KiRoEA90,KiRoEA93} or mixtures
\cite{RoDi00}. However, it does not predict freezing, which is actually
observed for the pure hard-sphere fluid at a packing fraction $\eta \simeq
0.494$. This deficiency can be resolved empirically by modifying the third
term in Eq.~(\ref{phi_RF}) \cite{RfEA96,RfEA97}, or more systematically by the
recipe of Tarazona \cite{Tar00} who introduced an additional tensorial
weighted density in the last term of $\Phi_{\mathrm{RF}}$. There remains
another factor which limits the accuracy of $\Phi_{\mathrm{RF}}$, namely the
equation of state obtained from Eq.~(\ref{eq_pTD}) or, equivalently,
Eq.~(\ref{eq_pSP}). One finds the pressure
\begin{equation}
\label{eq_eospy}
 \beta p_{\mathrm{PY}} = \frac{n_0}{1-n_3} + \frac{n_1 n_2}{(1-n_3)^2} +
  \frac{n_2^3}{12\pi(1-n_3)^3} \, ,
\end{equation}
which is the compressibility expression from the solution of the PY
integral equation \cite{Leb64}. The PY pressure is in good
agreement with simulations for the pure hard-sphere fluid at low packing
fractions but close to the freezing transition it overestimates the pressure
by about $7\%$. This problem can be solved by incorporating more accurate
equations of state within the context of Rosenfeld's FMT as an extrapolation
from low to high densities \cite{RoEvLaKa02,YuWu02}. In the next section we
introduce a new contribution along these lines.

\section{The new functional}
\label{sec_wbII}

An empirical correction of the high-density behavior of the PY
compressibility result has been given by Carnahan and Starling
\cite{CaSt69}. The CS equation of state has subsequently been generalized to
hard-sphere mixtures resulting in the
Boubl\'ik-Mansoori-Carnahan-Starling-Leland (BMCSL) pressure
$p_{\mathrm{BMCSL}}$ \cite{Bou70,MaCaStLe71}. This equation of state can be
written in terms of the weighted densities of a homogeneous hard-sphere
mixture:
\begin{equation}
  \beta p_{\mathrm{BMCSL}} =  \frac{n_0}{1-n_3} + \frac{n_1 n_2}{(1-n_3)^2} +
  \frac{n_2^3\left(1-\frac{1}{3}n_3\right)}{12\pi(1-n_3)^3} \, .
\end{equation}

Using this fact, $p_{\mathrm{BMCSL}}$ has been incorporated into FMT
\cite{RoEvLaKa02,YuWu02}. This was achieved by solving the differential
equation for the excess free energy $\Phi$ which is obtained by equating
$p_{\mathrm{BMCSL}}$ and the thermodynamic expression $p_{\mathrm{TD}}$ as
given in Eq.~(\ref{eq_pTD}). Note that for the implementation of this approach
a bulk fluid mixture is considered for which the vectorial weighted densities
vanish. Hence, the solution of the resulting differential equation is obtained
by using the dimensional ansatz Eq.~(\ref{eq_rfansatz}) without the vectorial
contributions. For this ansatz there is a unique solution if two additional
requirements are made: (i) the result for $\Phi$ is compatible with the
low-density limit Eq.~(\ref{eq_ldlimitwd}) and (ii) for the pure hard-sphere
fluid the third virial coefficient is recovered. Unlike in Rosenfeld's
derivation of $\Phi_{\mathrm{RF}}$ the vectorial contributions have to be
incorporated at a later stage. In analogy to $\Phi_{\mathrm{RF}}$ the
substitutions $n_1 n_2 \to n_1 n_2 - \bvec{n}_1 \cdot \bvec{n}_2$ and $n_2^3
\to n_2^3 - 3 n_2 \bvec{n}_2 \cdot \bvec{n}_2$ are made in $\Phi$. The
resulting functional $\Phi_{\mathrm{WB}}$ is called the White-Bear version of
FMT \cite{RoEvLaKa02}. In virtue of these substitutions, $\Phi_{\mathrm{WB}}$
has the correct low-density limit Eq.~(\ref{eq_ldlimitwd}) and the regularity
of the pair direct correlation function for $r\to0$ is guaranteed.

The White-Bear version of FMT has been shown to inherit all the good
properties of Rosenfeld's FMT for the description of the hard-sphere fluid and
improves the predictions of thermodynamic quantities due to the more accurate
underlying equation of state. This becomes particularly apparent in the contact
densities at a hard wall which are related to the pressure via a sum rule (for
a comparison with simulation data see, e.g.\ \cite{KoeBrMeRo05}). Furthermore,
one finds that the prediction of the freezing transition of the pure
hard-sphere system agrees very well with simulations \cite{RoEvLaKa02}. A
drawback of $\Phi_{\mathrm{WB}}$ is, however, that the scaled particle relation
Eq.~(\ref{eq_pSP}) is violated, i.e.\ one finds that $\partial
\Phi_{\mathrm{WB}} / \partial n_3 \ne p_{\mathrm{BMCSL}}$. This is of course
not surprising as the equality $p_{\mathrm{TD}}=p_{\mathrm{SP}}$ unambiguously
leads to Rosenfeld's $\Phi_{\mathrm{RF}}$, if we assume that the free energy
density is a function of the weighted densities $n_0,\dots,n_3$ and
$\bvec{n}_1$, and $\bvec{n}_2$ alone. Despite this inconsistency of the
White-Bear version, the quality of the resulting density distributions is high
\cite{BrRoScDi03}. However, analytical results obtained from the free energy
density $\Phi_{\mathrm{WB}}$ within the context of morphological
thermodynamics (\cite{KoeRoMe04} and references therein) are affected.

We conclude, that there is some room for improvement with respect to the
self-consistency of the free energy density. The basis for this improvement is
a new generalization of the CS pressure to mixtures of hard spheres which was
recently suggested by the authors \cite{HaRo06}. In terms of the weighted
densities, the new equation of state reads
\begin{equation}
\label{eq_CSIIIeos}
  \beta p_{\mathrm{CSIII}} =  \frac{n_0}{1-n_3} + \frac{n_1 n_2
  \left(1+\frac{1}{3} n_3^2  \right)}{(1-n_3)^2} + \frac{n_2^3
  \left(1-\frac{2}{3} n_3 + \frac{1}{3} n_3^2 \right)}{12\pi(1-n_3)^3} \, .
\end{equation}
The index CSIII refers to a hierarchy of extensions introduced in
Ref.~\cite{HaRo06}, where we showed for binary and ternary hard-sphere
mixtures that $p_{\mathrm{CSIII}}$ improves upon $p_{\mathrm{BMCSL}}$ compared
to computer simulations. Even more interestingly for the present context,
$p_{\mathrm{CSIII}}$ was constructed such that it is consistent with the
scaled-particle relation Eq.~(\ref{eq_pSP}) in the case of the one-component
hard-sphere fluid. Note, that consistency for the general hard-sphere
mixture within the framework of FMT would always result in the less accurate
pressure $p_{\mathrm{PY}}$.

By following the recipe for the derivation of the original White-Bear version
\cite{RoEvLaKa02,YuWu02}, described above, we calculate a new functional based
on the pressure $p_{\mathrm{CSIII}}$ 
\begin{eqnarray}
   \Phi_{\mathrm{WBII}} &=& - n_0 \ln (1-n_3) +
   {\textstyle \left(1+\frac{1}{9}n_3^2\phi_2(n_3)\right)}\frac{n_1 n_2 -
   \bvec{n}_1 \cdot \bvec{n}_2}{1-n_3}  \nonumber \\
  & & + \left(1-{\textstyle\frac{4}{9}}n_3\phi_3(n_3)\right)\frac{n_2^3 - 3
   n_2 \bvec{n}_2 \cdot \bvec{n}_2}{24\pi(1-n_3)^2}
  \label{eq_PhiWBII}
\end{eqnarray}
with
\begin{equation}
\hspace{-2.0cm}
\eqalign{
  \phi_2(n_3) & =  \left( 6 n_3 - 3 n_3^2 + 6 (1-n_3) \ln (1-n_3) \right) \big
  /n_3^3 = 1 + {\textstyle \frac{1}{2}} n_3 + \mathcal{O}(n_3^2) \, , \\
  \phi_3(n_3) & =  \left( 6 n_3 - 9 n_3^2 + 6 n_3^3 + 6 (1-n_3)^2 \ln(1-n_3)
  \right) \big/ (4n_3^3) = 1 -  {\textstyle \frac{1}{8}} n_3 +
  \mathcal{O}(n_3^2) \, .
}
\end{equation}
The new functional is an improvement of the White Bear version of FMT, as we
shall show in Sec.~\ref{sec_mtd}. The index WBII 
is chosen to indicate that the new functional is Mark II of the White Bear
functional.

For comparison we mention that in the above notation the original White-Bear
functional $\Phi_{\mathrm{WB}}$ is recovered with
$\phi_2^{\mathrm{WB}}(n_3)\equiv 0$ and
\begin{equation}
\hspace{-1.5cm}
  \phi_3^{\mathrm{WB}}(n_3) = \left( 9 n_3^2 - 6 n_3 - 6 (1-n_3)^2 \ln(1-n_3)
  \right) \big/ (4n_3^3) =  {\textstyle \frac{1}{2}} +  {\textstyle
  \frac{1}{8}} n_3 + \mathcal{O}(n_3^2) \, .
\end{equation}

We have compared predictions of our new version of FMT with corresponding
results obtained by the original White-Bear version for a pure hard-sphere
fluid and a binary mixture close to a planar hard wall. We have found that the
density distributions resulting from numerical minimization of the functional 
Eq.~(\ref{eq_gpfunc}) with $\Phi_{\mathrm{WB}}$ or $\Phi_{\mathrm{WBII}}$,
respectively, differ very little. For the pure hard-sphere fluid, this can be 
expected from the fact that the underlying bulk equation of state is the same
for both versions of FMT and hence the contact densities at the wall have to
be identical. Comparison with density distributions from Monte-Carlo
simulations revealed that the very small deviations of the DFT results from
the simulation data are clearly more significant than the mutual deviations
between the two FMT versions. We conclude that the limitations of FMT-based
density functionals cannot be considerable pushed forward by increasing the
quality of the underlying bulk equation of state but are rather determined by
the structure of FMT itself, i.e.\ the set of weight functions which are
employed and hence the restriction to one-center convolutions. For a
discussion of this topic see Ref.~\cite{CuMaTa02}. A slight improvement from 
the WBII version is indeed found for the description of the pair direct
correlation function as can be inferred from comparison with simulation data
(not shown).

We find the main benefit of the new functional $\Phi_{\mathrm{WBII}}$ in the
context of morphological thermodynamics. Here, the self-consistency of
$\Phi_{\mathrm{WBII}}$ on the level of the pressure, i.e.\ the equality of
$p_{\mathrm{TD}}$ from Eq.~(\ref{eq_pTD}) and $p_{\mathrm{SP}}$ from
Eq.~(\ref{eq_pSP}) in the case of the pure fluid, is crucial for the accuracy
of analytical expressions obtained within the morphological theory. In the
next section we give a brief introduction to the theory and show examples
which illustrate the gain from the new functional $\Phi_{\mathrm{WBII}}$. 

\section{Morphological thermodynamics}
\label{sec_mtd}

The morphometric approach to the grand potential of a fluid around a complexly
shaped object $\mathcal{B}$ (or a fluid inside a complexly shaped container) was
inspired by the Hadwiger theorem \cite{Had57} from integral geometry
\cite{Me94}. The theorem states that every motion-invariant, continuous and
additive functional of the complexly shaped object $\mathcal{B}$ depends on
the shape of $\mathcal{B}$ via only four geometric measures: the volume $V$,
the surface area $A$, the integrated (over the surface area) mean and Gaussian
curvature $C$ and $X$, respectively. The latter are given as
\begin{equation}
C = \int\limits_{\partial \mathcal{B}}\!\upd \bvec{r}\, \frac{1}{2} \left( \frac{1}{R_1} +
  \frac{1}{R_2} \right)
\, , \quad
X = \int\limits_{\partial \mathcal{B}} \! \upd \bvec{r}\,\frac{1}{R_1 R_2} \, .
\end{equation}
Here $R_1$ and $R_2$ are the local principal radii of curvature on the surface
of $\mathcal{B}$. Note that $X$ is proportional to the Euler characteristic.

While the Hadwiger theorem is rigorous, the connection to physics is not
obvious and cannot be proven rigorously. However, there is strong
numerical evidence
\cite{BrRoMeDi03,KoeRoMe04,KoeBrMeRo05,Ro05,RoHaKi06} that the solvation free
energy of a convex body immersed in a solvent away from the critical point and
away from wetting or drying transitions takes the form 
\begin{equation}
\label{eq_dommorpho}
 \Delta \Omega = p V + \sigma A + \kappa C + \bar{\kappa} X \, .
\end{equation}
where the conjugated quantities to the geometric measures of $\mathcal{B}$ are
thermodynamic coefficients depending only on the temperature, the chemical
potentials and the given interaction between $\mathcal{B}$ and the fluid, and
among fluid particles, but not on the (complex) geometry of $\mathcal{B}$. 
The thermodynamic coefficients are $p$, the pressure, $\sigma$ the planar wall
surface tension, and $\kappa$ and $\bar{\kappa}$ two bending rigidities.

Morphometry is obviously a very useful tool for the calculation of
thermodynamic quantities in complex geometries as it allows one to calculate 
the shape-independent thermodynamic coefficients in simple geometry. The
treatment of the actual complex geometry (or a set of different geometries)
then only requires a straightforward calculation of the geometric
measures. For instance, the morphometric approach has been applied
successfully to the calculation of solvation free energies of a protein in
various geometrical configurations \cite{RoHaKi06} and to the thermodynamics 
of fluids in porous media \cite{MeAr05}. 

The test geometry we choose here is the case where the particle $\mathcal{B}$
is a single hard sphere $\mathcal{S}$ of radius $R_{\mathrm{s}}$ immersed
in a pure hard-sphere fluid with radius $R$ and density $\rho$. The change in
grand potential $\Delta\Omega$ due to the insertion of the sphere
$\mathcal{S}$ is obtained by minimizing the density functional
Eq.~(\ref{eq_gpfunc}) with either $\Phi_{\mathrm{WB}}$ or
$\Phi_{\mathrm{WBII}}$. From the equilibrium density profile
$\rho_0(\bvec{r})$ one can calculate $\Delta  \Omega =
\Omega[\rho_0(\bvec{r})]-\Omega[\rho(\bvec{r})=\rho_{\mathrm{bulk}}]$. By 
repeating the calculation for different values of $R_{\mathrm{s}}$ the
function $\Delta \Omega(R_{\mathrm{s}})$ is obtained numerically.

On the other hand, we have the morphometric prediction for  $\Delta \Omega$,
Eq.~(\ref{eq_dommorpho}). In order to evaluate the morphometric solvation free
energy it is most convenient to calculate the geometrical measures and the
thermodynamic coefficients at the surface at which the density profile
$\rho_0(\bvec{r})$ jumps discontinuously to zero. This surface is parallel to
the physical wall of $\mathcal{S}$ at normal distance $R$. Note that it is
actually the parallel surface and not the physical surface that enters 
in the external potential $V_{\mathrm{ext}}(r)$ in Eq.~(\ref{eq_gpfunc}) that
$\mathcal{S}$ exerts on the fluid. In terms of the parallel surface, which is
simply a sphere with radius $R_{\mathrm{s}}+R$, the morphometric form,
Eq.~(\ref{eq_dommorpho}), reads 
\begin{equation}
\label{eq_domsphere}
\hspace{-1.0cm}
  \Delta\Omega(R_{\mathrm{s}}) =  p \, \frac{4}{3}\pi(R_{\mathrm{s}}+R)^3 +
  \sigma \, 4\pi(R_{\mathrm{s}}+R)^2 + \kappa \, 4\pi(R_{\mathrm{s}}+R) +
  \bar{\kappa} \, 4\pi \, . 
\end{equation}

The extraction of the thermodynamic coefficients $p$, $\sigma$, $\kappa$ and
$\bar{\kappa}$ from the values $\Delta\Omega(R_{\mathrm{s}})$ obtained by
minimization of the density functional is therefore achieved by fitting
Eq.~(\ref{eq_domsphere}) to the numerical DFT data for different values of
$R_{\mathrm{s}}$. This fit was performed for the data from
$\Phi_{\mathrm{WB}}$ and $\Phi_{\mathrm{WBII}}$ in the range
$R_{\mathrm{s}}\in [2R,10R]$ for various values of the packing fractions of the
fluid. Indeed, we find the assumption made by Eq.~(\ref{eq_domsphere}) on the 
$R_{\mathrm{s}}$-dependence of $\Delta\Omega$ clearly confirmed and in
accordance with previous results \cite{KoeRoMe04}. We shall come back to our
results later on, referring to them as obtained via the ``minimization route''.

In virtue of its applicability to mixtures, FMT also provides analytical
expressions for the thermodynamic coefficients $p$, $\sigma$, $\kappa$ and
$\bar{\kappa}$. For their derivation, we follow the ideas of
Refs.~\cite{BrRoMeDi03,OvRo05}. In Ref.~\cite{BrRoMeDi03} the curvature
dependence of the excess surface grand potential and contact density of a
hard-sphere fluid in contact with hard curved walls was studied. The basic
idea is to consider a binary bulk mixture consisting of a hard-sphere fluid
with radius $R$ and packing fraction $\eta$ and a single sphere
$\mathcal{S}$, which is the second component at infinite dilution, i.e.\
$\rho_{\mathrm{s}}\to 0$. $\Delta\Omega$ is then obtained as the
excess chemical potential $\mu_{\mathrm{s}}^{\mathrm{ex}}$.
$\mu_{\mathrm{s}}^{\mathrm{ex}}$ can be calculated as the derivative of the
mixture excess free energy density with respect to $\rho_{\mathrm{s}}$. Using
any of the above FMT expressions for the excess free energy $\Phi$ we find
\begin{eqnarray}
  \beta\Delta\Omega & = & \beta \mu_{\mathrm{s}}^{\mathrm{ex}} =
  \lim_{\rho_{\mathrm{s}}\to 0}\frac{\partial
  \Phi }{\partial \rho_{\mathrm{s}}} \nonumber \\
  & = & \frac{\partial\Phi }{\partial n_3} \,\frac{4}{3}\pi R_{\mathrm{s}}^3  +
   \frac{\partial\Phi }{\partial n_2} \, 4\pi R_{\mathrm{s}}^2 +
   \frac{\partial\Phi }{\partial n_1} \, R_{\mathrm{s}} +
   \frac{\partial\Phi }{\partial n_0} \, .
   \label{eq_morphSphere}
\end{eqnarray}  

Note that all vectorial contributions in $\Phi$ vanish in the uniform
bulk. Due to the limit $\rho_{\mathrm{s}} \to 0$ the
partial derivatives of $\Phi$ are evaluated for the solvent only, i.e.\ a
one-component uniform fluid with radius $R$ and packing fraction $\eta$.

A comparison of Eqs.~(\ref{eq_domsphere}) and (\ref{eq_morphSphere}) allows to
identify the thermodynamic coefficients, calculated for the parallel surface,
with certain linear combinations of the partial derivatives of $\Phi$. We find
that
\begin{equation}
\eqalign{
  \beta p & =  \frac{\partial \Phi}{\partial n_3} \, , \\
  \beta \sigma & =  \frac{\partial \Phi}{\partial n_2} 
                 - R \frac{\partial \Phi}{\partial n_3} \, , \\
  \beta \kappa & =  \frac{1}{4\pi} \frac{\partial \Phi}{\partial n_1}
                 - 2 R \frac{\partial \Phi}{\partial n_2}
                 + R^2 \frac{\partial \Phi}{\partial n_3} \, ,\\
  \beta \bar{\kappa} & =  \frac{1}{4\pi} \frac{\partial \Phi}{\partial n_0}
                 - \frac{R}{4\pi} \frac{\partial \Phi}{\partial n_1}
                 + R^2 \frac{\partial \Phi}{\partial n_2}
                 - \frac{1}{3} R^3 \frac{\partial \Phi}{\partial n_3} \, .
}
\end{equation}

The relation for the pressure is precisely the scaled particle relation,
Eq.~(\ref{eq_pSP}). In the following, we refer to the above analytical results
for the thermodynamic coefficients as the outcome of the ``bulk route''.

We give the explicit results for the coefficients only for the case of the
new excess free energy density $\Phi_{\mathrm{WBII}}$:
\begin{equation}\label{eq_coeffHSWBII}
\eqalign{
  \frac{\beta p_{\mathrm{WBII}}}{\rho} & = 
  \frac{1+\eta+\eta^2-\eta^3}{(1-\eta)^3} \, , \\
  \frac{\beta \sigma_{\mathrm{WBII}}}{R\rho} & =  -
  \frac{1+2\eta+8\eta^2-5\eta^3}{3(1-\eta)^3} - \frac{\ln(1-\eta)}{3\eta} \, ,
  \\
  \frac{\beta \kappa_{\mathrm{WBII}}}{R^2\rho} & =  
  \frac{4-10\eta+20\eta^2-8\eta^3}{3(1-\eta)^3} + \frac{4\ln(1-\eta)}{3\eta}
  \, , \\
  \frac{\beta \bar{\kappa}_{\mathrm{WBII}}}{R^3\rho} & = 
  \frac{-4+11\eta-13\eta^2+4\eta^3}{3(1-\eta)^3} - \frac{4\ln(1-\eta)}{3\eta}
  \, .
}
\end{equation}

We emphasize that the pressure $p_{\mathrm{WBII}}$ is precisely the
quasi-exact CS expression. This is not a trivial fact but
rather a consequence of the construction of the novel mixture equation of state
Eq.~(\ref{eq_CSIIIeos}) \cite{HaRo06}. In contrast, the original White-Bear
version of FMT, which is based on a different mixture generalization of the
CS pressure \cite{Bou70,MaCaStLe71} does not possess this
feature of self-consistency, i.e.\ the derivative of $\Phi_{\mathrm{WB}}$ with
respect to $n_3$ does {\em not} yield exactly the CS pressure
\cite{RoEvLaKa02}.

\begin{figure}[tbp]
  \psfrag{c1}{\large$p$}
  \psfrag{c2}{\large$\sigma$}
  \psfrag{c3}{\large$\kappa$}
  \psfrag{c4}{\large$\bar{\kappa}$}
  \psfrag{tdcoeff}{\hspace{-1cm}\large $\beta p R^3$, $\beta \sigma R^2$, $\beta \kappa R$,
    $\beta \bar{\kappa}$}
  \psfrag{eta}{\large $\eta$}
  
  \vspace*{1.0cm}

  \begin{center}
    \includegraphics[width = 13cm]{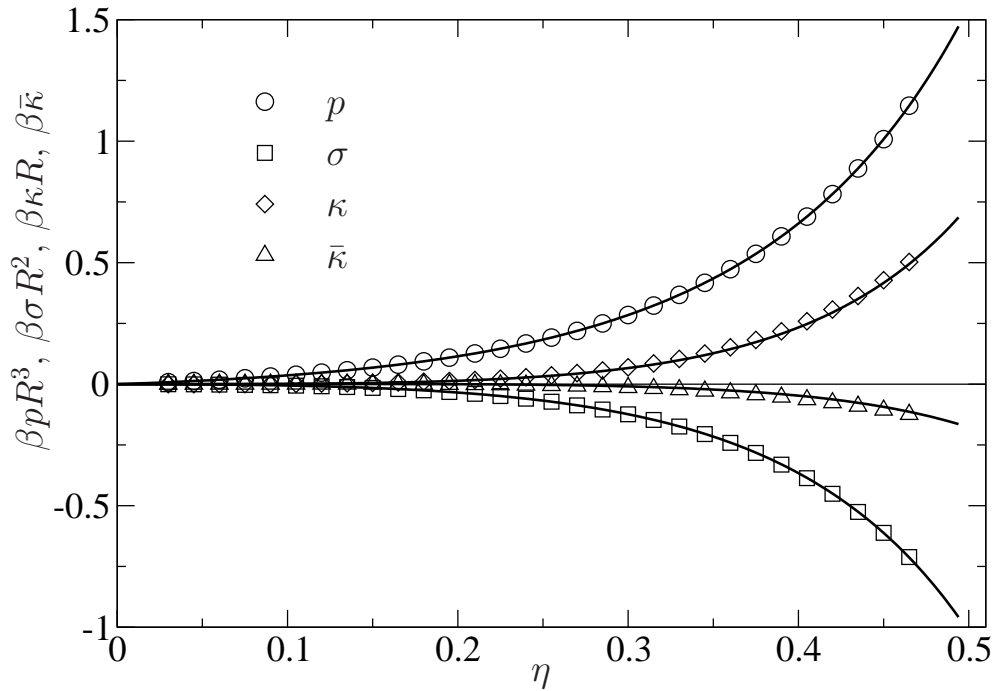}
  \end{center}

  \begin{center}
    \caption{Results for the four thermodynamic coefficients $p$, $\sigma$,
      $\kappa$, and $\bar \kappa$ of the hard-sphere fluid are shown as
      obtained from the new excess free energy density $\Phi_{\textrm{WBII}}$,
      Eq.~(\ref{eq_PhiWBII}). The analytical expressions given in
      Eqs.~(\ref{eq_coeffHSWBII}) are denoted by the lines, while the results
      from the minimization route are given by the symbols. $\eta$ is the
      packing fraction. At $\eta\approx 0.494$ the hard-sphere fluid freezes.} 
\label{fig_coeffHS}
  \end{center}
\end{figure}
   
We now compare the results from the bulk and minimization routes as obtained
for the different versions $\Phi_{\mathrm{WB}}$ and $\Phi_{\mathrm{WBII}}$
of FMT. In Fig.~\ref{fig_coeffHS} we show our results for the thermodynamic
coefficients calculated from the new functional $\Phi_{\mathrm{WBII}}$. The
agreement for $p$ is perfect by construction of the equation of state,
Eq.~(\ref{eq_CSIIIeos}), and very good for the surface tension $\sigma$. Note
that from a comparison with simulation data $\sigma_{\mathrm{WBII}}$ was shown
previously to be of high accuracy for intermediate and high packing
fractions of the hard-sphere solvent. At low packing fractions, however, we
found a small deviation from the exact low density limit of $\sigma$
\cite{HaRo06}. Only for the bending rigidities $\kappa$ and $\bar{\kappa}$ a
slight inconsistency between the bulk and the minimization route
appears. However, this inconsistency remains below $1\%$ at high values of
$\eta$ and we conjecture from the very good agreement of
$\sigma_{\mathrm{WBII}}$ with simulation data that also
$\kappa_{\mathrm{WBII}}$ and $\bar{\kappa}_{\mathrm{WBII}}$ deliver accurate 
expressions for the hard-sphere fluid thermodynamic coefficients. 
With Eqs.~(\ref{eq_coeffHSWBII}) we have obtained a set of analytical
expressions for the thermodynamic coefficients that are more accurate than
previous suggestions, namely the results calculated from the original
White-Bear version or those from Rosenfeld's DFT.

\begin{figure}[tbp]
  \psfrag{tdcoeff1}{\large$\beta \Delta p R^3$}
  \psfrag{tdcoeff2}{\large$\beta \Delta \sigma R^2$}
  \psfrag{tdcoeff3}{\large$\beta \Delta \kappa R$}
  \psfrag{tdcoeff4}{\hspace{0.2cm}\large$\beta \Delta \bar{\kappa}$}
  \psfrag{hlab1}{\large$p - p_{\mathrm{WBII}}$}
  \psfrag{hlab2}{\large$\sigma - \sigma_{\mathrm{WBII}}$}
  \psfrag{hlab3}{\large$\kappa - \kappa_{\mathrm{WBII}}$}
  \psfrag{hlab4}{\large$\bar{\kappa} - \bar{\kappa}_{\mathrm{WBII}}$}
  \psfrag{eta}{\large $\eta$}
  
  \vspace*{1.0cm}

  \begin{center}
    \includegraphics[width = 13cm]{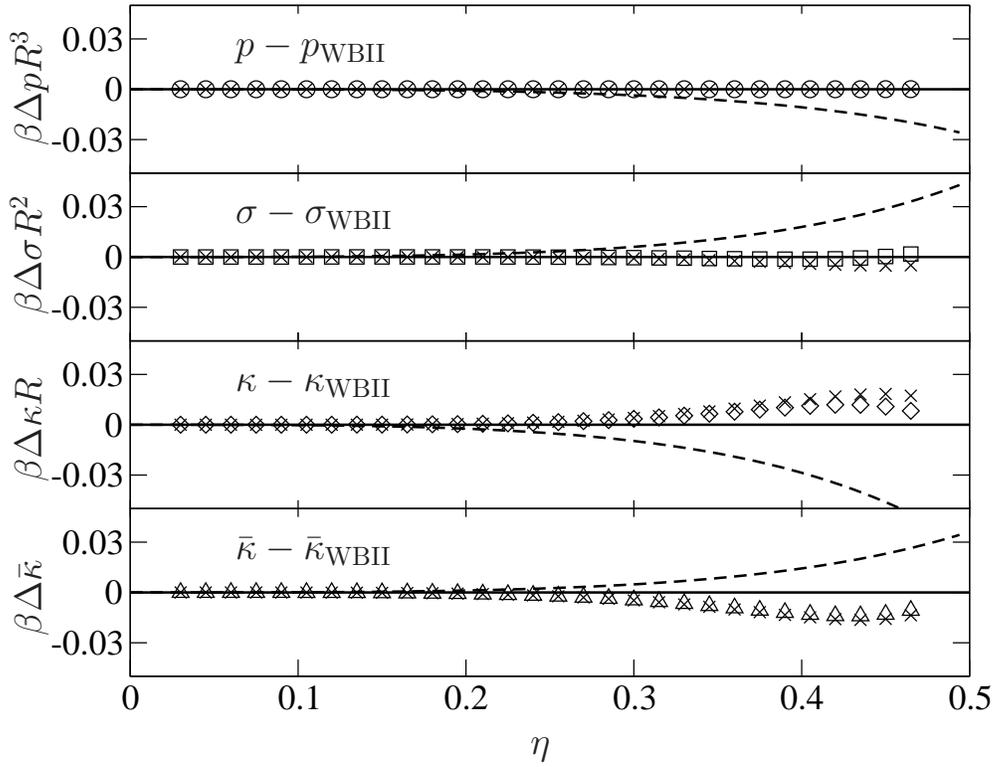}
  \end{center}
  
  \begin{center}
    \caption{Various results for the four thermodynamic coefficients, $p$,
      $\sigma$, $\kappa$, and $\bar \kappa$. Shown are the differences of
      these coefficients obtained by various routes and theories to the
      analytical WBII results, Eqs.~(\ref{eq_coeffHSWBII}), cf.\ the lines in
      Fig.~\ref{fig_coeffHS}. The symbols (except the crosses) denote the WBII
      results from the minimization route. For comparison, we also show
      results from the original White-Bear version: those from the bulk route
      are plotted as dashed lines while the crosses denote the outcome of the
      minimization route.}
    \label{fig_compcoeff}
  \end{center}
\end{figure}

As an illustration, we plot in Fig.~\ref{fig_compcoeff} the difference of
various results for the four thermodynamic coefficients, $p$, $\sigma$,
$\kappa$, and $\bar \kappa$ from the analytical expressions 
Eqs.~(\ref{eq_coeffHSWBII}) of the WBII version. Again, we find a high
degree of self-consistency of the new functional $\Phi_{\mathrm{WBII}}$
(symbols, except the crosses in Fig.~\ref{fig_compcoeff}). In contrast,
the inconsistency of the original White-Bear version, which can be seen by the
distance between the dashed line and the crosses in Fig.~\ref{fig_compcoeff}, is considerably larger and appears even
for the pressure. The analytical expressions derived in the bulk route from
$\Phi_{\mathrm{WB}}$ are therefore of a lower quality than
Eqs.~(\ref{eq_coeffHSWBII}), which manifests itself also in their poorer
agreement with simulations \cite{HaRo06}. The good agreement between the
results from the minimization route for the two versions of FMT is a direct
consequence of the good agreement in the corresponding density profiles. This
observation can be rationalized by noting that the contact value of the
density profile at a planar wall coincides for both version of FMT as a result
of the same bulk equation of state.

We do not include the results from Rosenfeld's functional $\Phi_{\mathrm{RF}}$
in Fig.~\ref{fig_compcoeff} as this would require to extend the range of the vertical axis
considerably and therefore obscure the examination of consistency of the
White-Bear versions. When calculated from $\Phi_{\mathrm{RF}}$, however, the pressure follows the PY compressibility result which
quantitatively differs from simulations so that the analytical expressions
from the bulk route only yield a qualitative description of the thermodynamic
coefficients. Surprisingly, the agreement between the bulk and the
minimization route is comparable to that of $\Phi_{\mathrm{WB}}$ except for
the pressure where $\Phi_{\mathrm{RF}}$ is consistent \cite{Koe05}. 
Intuitively, one might expect a better agreement for $\Phi_{\mathrm{RF}}$ than
for $\Phi_{\mathrm{WBII}}$ because of the self-consistency on the level of
the pressure. This interesting feature will be encountered again in the
following where we consider the contact density at a curved hard wall.

The contact density of the hard-sphere fluid at a hard wall is connected to
the normal derivative of the grand potential $\Omega$ \cite{KoeRoMe04}. The
case of interest here is again a hard-sphere fluid (radius $R$, packing
fraction $\eta$) around a sphere $\mathcal{S}$ with radius
$R_{\mathrm{s}}$. We determine the grand potential from the density functional
$\Omega[\rho_0(\bvec{r})]$. The normal derivative of $\Omega$ reduces due to
the symmetry to a derivative with respect to $R_{\mathrm{s}}$ at constant
chemical potential, which is then calculated as 
\begin{equation}
 \frac{\partial\Omega}{\partial R_{\mathrm{s}}} = \int\!\upd
 \bvec{r}\,\frac{\delta
 \Omega[\rho_0(\bvec{r})]}{\delta \rho} \frac{\partial
 \rho_0(\bvec{r})}{\partial R_{\mathrm{s}}} + \int\!\upd \bvec{r}\,
 \rho_0(\bvec{r}) \frac{\partial V_{\mathrm{ext}}(\bvec{r})}{\partial
 R_{\mathrm{s}}} \, .
\end{equation}
The first integral vanishes due to the equilibrium condition for
$\rho_0(\bvec{r})$, i.e.\ $\delta \Omega / \delta \rho = 0$. The derivative of
the external potential gives rise to a $\delta$-peak at the location of the
parallel wall, and one finds \cite{He83} 
\begin{equation}
\label{eq_gencontdens}
 \beta \, \frac{\partial \Omega}{\partial R_{\mathrm{s}}} = 4 \pi
 (R_{\mathrm{s}}+R)^2 \rho_{\mathrm{c}}
\end{equation} 
where $\rho_{\mathrm{c}}$ is the contact value of the density of the fluid at
the sphere $\mathcal{S}$. Using the morphometric form
Eq.~(\ref{eq_domsphere}) for $\Delta\Omega$, the grand potential $\Omega$ of
the fluid containing the sphere $\mathcal{S}$ is
$\Omega(R_{\mathrm{s}}) = -p V_{\mathrm{tot}}+\Delta\Omega(R_{\mathrm{s}})$,
where $V_{\mathrm{tot}}$ is the total volume of the system. In the
thermodynamic limit $V_{\mathrm{tot}}\to \infty$. If the morphometric form is
inserted into Eq.~(\ref{eq_gencontdens}) one obtains the contact density
$\rho_{\mathrm{c}}$ 
\begin{equation}
\label{eq_morphrhocon}
  \rho_{\mathrm{c}} = \beta p + \frac{2 \beta \sigma}{R_{\mathrm{s}}+R} +
  \frac{\beta \kappa}{(R_{\mathrm{s}}+R)^2} \, .
\end{equation}
For $R_{\mathrm{s}}\to \infty$ the planar wall contact theorem
$\rho_{\mathrm{c}} = \beta p$ is recovered and for finite values of
$R_{\mathrm{s}}$ the contact density is lowered.

\begin{figure}[tbp]
  \psfrag{rf}{\hspace{-0.4cm}\large DFT $\Phi_{\mathrm{RF}}$}
  \psfrag{wb}{\hspace{-0.4cm}\large DFT $\Phi_{\mathrm{WB}}$}
  \psfrag{csIII}{\hspace{-0.4cm}\large DFT $\Phi_{\mathrm{WBII}}$}
  \psfrag{rfmorph}{\hspace{-0.2cm}\large RF Eq.~(\ref{eq_morphrhocon})}
  \psfrag{wbmorph}{\hspace{-0.2cm}\large WB Eq.~(\ref{eq_morphrhocon})}
  \psfrag{csIIImorph}{\hspace{-0.2cm}\large WBII Eq.~(\ref{eq_morphrhocon})}
  \psfrag{R/(Rs+R)}{\large$R/(R_{\mathrm{s}}+R)$}
  \psfrag{rhocont}{\large$\rho_{\mathrm{c}} R^3$}
  \psfrag{devrhoc}{\large$\Delta\rho_{\mathrm{c}} R^3$}
  \psfrag{devlabel}{\large$\rho_{\mathrm{c}}-(\rho_{\mathrm{c}})_{\mathrm{WBII}}$}

  \vspace*{1.0cm}

  \begin{center}
    \includegraphics[width = 13cm]{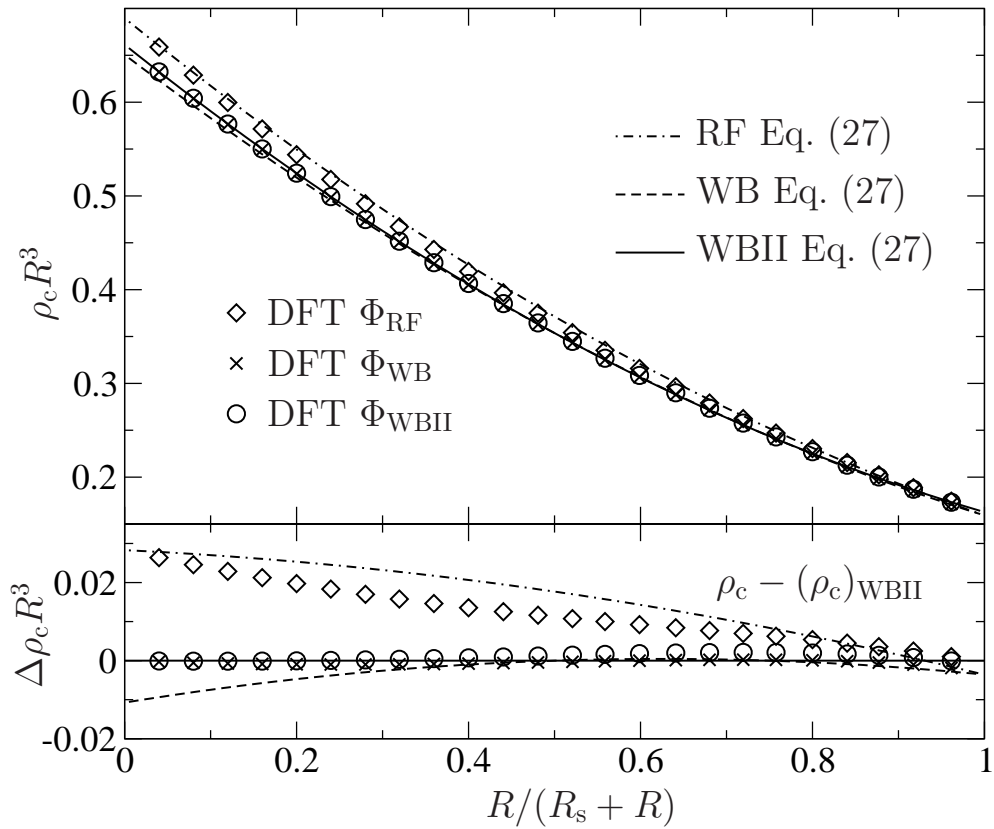}
  \end{center}
  
  \begin{center}
    \caption{Contact value $\rho_{\mathrm{c}}$ of the density of a
      hard-sphere fluid (radius $R$, packing fraction $\eta=0.4$) at a single
      sphere with radius $R_{\mathrm{s}}$. We show results obtained from the
      excess free energy densities $\Phi_{\mathrm{RF}}$, $\Phi_{\mathrm{WB}}$
      and $\Phi_{\mathrm{WBII}}$, respectively. We compare results from the
      numerical minimization of the density functional (symbols),
      Eq.~(\ref{eq_gpfunc}), with the morphometric prediction (lines) according
      to Eq.~(\ref{eq_morphrhocon}).} 
\label{fig_contden}
  \end{center}
\end{figure}

We show results for the contact density $\rho_{\mathrm{c}}$ of a hard-sphere
fluid with packing fraction $\eta = 0.4$ as a function of the inverse of the
radius $R_{\mathrm{s}}$ in Fig.~\ref{fig_contden}. The symbols are the contact
densities obtained from the density profiles which we calculated by minimizing
numerically the functional Eq.~(\ref{eq_gpfunc}). The lines in
Fig.~\ref{fig_contden} show the morphometric prediction according to
Eq.~(\ref{eq_morphrhocon}) with the analytical expressions for the
thermodynamic coefficients from the different versions of  FMT (cf.\ the bulk
route above). The first observation we make is that the numerical results from
$\Phi_{\mathrm{WB}}$ and $\Phi_{\mathrm{WBII}}$ are nearly indistinguishable
and indeed in the planar wall limit, $R_{\mathrm{s}}\to\infty$, the data 
coincide by construction of the functionals. Taking this fact into account it
is understandable that the results for $\rho_{\mathrm{c}}$ for finite values of
$R_{\mathrm{s}}$ are very similar. The numerical data for $\Phi_{\mathrm{RF}}$
tends towards the PY pressure for $R_{\mathrm{s}}\to\infty$ which is known to 
overestimate the actual pressure in the hard-sphere fluid for sufficiently
high values of $\eta$. In the limit $R_{\mathrm{s}}\to 0$ (point-like object)
the data from the three versions of FMT coincide.

Comparing with the analytical prediction from morphometry, we find very good
agreement between the results from $\Phi_{\mathrm{WBII}}$ over the whole
interval of $R_{\mathrm{s}}$. Only for very small radii $R_{\mathrm{s}}$ a
slight deviation is visible. Therefore, the new functional improves upon the
results of the original White Bear version of FMT, which performs well at
small radii, but produces a small error at large values of $R_{\mathrm{s}}$
due to the the inconsistency of $\Phi_{\mathrm{WB}}$ in the pressure. 

As mentioned above for the thermodynamic coefficients, a moderate agreement in
the case of Rosenfeld's FMT is also observed for the contact density. While
the approach is consistent for large values of $R_{\mathrm{s}}$ by
construction of the functional, in the range of smaller values of
$R_{\mathrm{s}}$ a deviation is clearly visible. This behavior has been
observed previously \cite{BrRoMeDi03}. We find this fact remarkable because it
shows that, from the point of view of self-consistency, the new hard-sphere
mixture equation of state  Eq.~(\ref{eq_CSIIIeos}) is better suited for an
implementation within FMT than the PY mixture equation of state
Eq.~(\ref{eq_eospy}) itself. This is even more surprising as the latter is
characterized by full consistency for mixtures on the level of the pressure. 

\section{Conclusion}

\label{sec_concl}

In this work, we have introduced a new density functional for hard-sphere
mixtures which, in the spirit of the original White-Bear version, incorporates
the quasi-exact CS equation of state within the framework of
FMT. While the original White-Bear version is based on the well-known BMCSL
equation of state, the White-Bear version Mark II, presented here, is derived
from an improved mixture generalization of the CS equation of state recently
introduced by us \cite{HaRo06}. The new functional WBII, besides having all
the good properties of the original White-Bear version, improves the level of
self-consistency. The level of consistency of the WBII version for
the pure hard-sphere fluid is examined in the context of morphological 
thermodynamics. Our study reveals that, beside the improved consistency of the
pressure, also the consistency of the surface tension $\sigma$ and the
bending rigidities $\kappa$ and $\bar \kappa$ is clearly improved. Supported
by a previous comparison to simulation data for the surface tension
\cite{HaRo06} we can argue that the thermodynamic quantities derived from the
new functional WBII are on a par with simulations.

We have presented evidence that in the case of the pure hard-sphere fluid the
degree of self-consistency of the WBII version is even higher than that of
Rosenfeld's original FMT. This is a remarkable finding as Rosenfeld's FMT is
by construction fully (i.e.\ for an arbitrary mixture) consistent on the level
of the pressure. Apparently, this fact does not translate into a high level
of consistency for other thermodynamic quantities (such as surface tension and
bending rigidities) of the pure fluid. We conclude that, in what concerns the
pure hard-sphere fluid, the recent mixture generalization of the CS equation
of state Eq.~(\ref{eq_CSIIIeos}) is even better suited as a staring
point for FMT than the PY compressibility mixture equation,
underlying Rosenfeld's FMT. Note further that the PY mixture
equation of state deviates significantly from simulations at sufficiently high
densities.

In conclusion, with the WBII version we have constructed a new hard-sphere
functional based on the CS pressure which improves upon the
original White-Bear functional. Although the differences between the density
profiles in simple geometries resulting from minimization of the functionals
are small, the increased self-consistency of the WBII version proves crucial
for analytical calculations within the context of morphological
thermodynamics.

Our considerations are based on thermodynamic arguments and result in a change
of the dependency of the free energy density $\Phi$ on the weighted density
$n_3$. There are several other developments in FMT that were mainly
concerned with improving the performance of FMT in highly confined
geometries. These studies suggest to change the dependency of $\Phi$ on $n_2$
and $\bvec{n}_2$ \cite{RfEA96,RfEA97} or to introduce new tensorial weighted
densities \cite{Tar00}. It is worth pointing out that these improvements
concerning the description of hard-sphere fluids in highly confined geometries
are straightforwardly combined with the improvements on thermodynamics
presented here.


\section*{References}
\begin{thebibliography}{99}

\bibitem{Ev79} Evans R (1979) Adv. Phys. {\bf 28} 143.
\bibitem{Ev90} Evans R (1990) J. Phys.: Condens. Matter {\bf 2} 8989.
\bibitem{Wu06} Wu J (2006) AIChE J {\bf 52} 1169.
\bibitem{Ev92} Evans R (1992) in {\it Fundamentals of Inhomogeneous Fluids},
(ed. D.~Henderson), 85 (Marcel Dekker, New York, NY)
\bibitem{HaMcDo} Hansen JP and McDonald IR, (1986) \textit{Theory of simple
    liquids} (Academic Press, London).
\bibitem{RuEA96} Rutgers MA, Dunsmuir JH, Xue J-Z, Russel WB, and
  Chaikin PM (1996) Phys. Rev. B {\bf 53} 5043.
\bibitem{RoEvDi00} Roth R, Evans R, and Dietrich S (2000) Phys. Rev. E {\bf
  62} 5360.
\bibitem{GrEA04} Grodon C, Dijkstra M, Evans R, and Roth R (2004)
  J. Chem. Phys. {\bf 121} 7869.
\bibitem{GrEA05} Grodon C, Dijkstra M, Evans R, and Roth R (2005)
  Mol. Phys. {\bf 103} 3009. 
\bibitem{Rf89} Rosenfeld Y (1989) Phys. Rev. Lett. {\bf 63} 980.
\bibitem{RoEvLaKa02} Roth R, Evans R, Lang A, and Kahl G (2002) J. Phys.:
Condens. Matter {\bf 14} 12063.
\bibitem{YuWu02} Yu Y-X and Wu J (2002) J. Chem. Phys. {\bf 117} 10156.
\bibitem{BrRoScDi03} Bryk P, Roth R, Schoen M, and Dietrich S (2003)
  Europhys. Lett. {\bf 63} 233.
\bibitem{HaRo06} Hansen-Goos H and Roth R (2006) J. Chem. Phys. {\bf 124}
  154506.
\bibitem{VaDaPe89} Vanderlick TK, Davis HT, and Percus JK (1989)
  J. Chem. Phys. {\bf 91} 7136.
\bibitem{Rf90} Rosenfeld Y (1990) Phys. Rev. A {\bf 42} 5978.
\bibitem{KiRoEA90} Kierlik E and Rosinberg ML (1990) Phys. Rev. A {\bf 42}
  3382.
\bibitem{KiRoEA93} Phan S, Kierlik E, Rosinberg ML, Bildstein B, and Kahl G
  (1993) Phys. Rev. E {\bf 48} 618.
\bibitem{RoDi00} Roth R and Dietrich S (2000) Phys. Rev. E {\bf 62} 6926.
\bibitem{RfEA96} Rosenfeld Y, Schmidt M, L\"owen H, and Tarazona P (1996)
  J. Phys.: Condens. Matter {\bf 8} L577.
\bibitem{RfEA97} Rosenfeld Y, Schmidt M, L\"owen H, and Tarazona P (1997)
 Phys. Rev. E {\bf 55} 4245.
\bibitem{Tar00} Tarazona P (2000) Phys. Rev. Lett. {\bf 84} 694.
\bibitem{Leb64} Lebowitz JL (1964) Phys. Rev. {\bf 133} A895.
\bibitem{CaSt69} Carnahan NF and Starling KE (1969) J. Chem. Phys. {\bf 51}
  635.
\bibitem{Bou70} Boubl\'ik T (1970) J. Chem. Phys. {\bf 53} 471.
\bibitem{MaCaStLe71} Mansoori GA, Carnahan NF, Starling KE, and
  Leland TW (1971) J. Chem. Phys. {\bf 54} 1523.
\bibitem{KoeBrMeRo05} K\"onig P-M, Bryk P, Mecke K, and Roth R (2005)
  Europhys. Lett. {\bf 69} 832.
\bibitem{KoeRoMe04} K\"onig P-M, Roth R and Mecke KR (2004)
  Phys. Rev. Lett. {\bf 93} 160601.
\bibitem{CuMaTa02} Cuesta JA, Mart\'inez-Rat\'on Y, and Tarazona P (2002)
  J. Phys.: Condens. Matter {\bf 14} 11965.
\bibitem{Had57} Hadwiger H (1957) {\it Vorlesungen \"uber Inhalt, Oberfl\"ache
    und Isoperimetrie} (Springer, Berlin, Germany).
\bibitem{Me94} Mecke KR (1994) {\it Integralgeometrie in der Statistischen
  Physik} (Harri Deutsch, Frankfurt, Germany).
\bibitem{BrRoMeDi03} Bryk P, Roth R, Mecke KR, and Dietrich S (2003)
  Phys. Rev. E {\bf 68} 031602.
\bibitem{Ro05} Roth R (2005) J. Phys.: Condens. Matter {\bf 17} S3463.
\bibitem{RoHaKi06} Roth R, Harano Y, and Kinoshita M (2006)
  Phys. Rev. Lett., {\it accepted}. 
\bibitem{MeAr05} Mecke K and Arns CH (2005) J. Phys.: Condens. Matter {\bf 17}
  S503.
\bibitem{OvRo05} Oversteegen SM and Roth R (2005) J. Chem. Phys. {\bf 122}
  214502. 
\bibitem{Koe05} K\"onig P-M (2005) doctoral thesis, ITAP, University of
  Stuttgart, Stuttgart.
\bibitem{He83} Henderson JR (1983) Mol. Phys. {\bf 50} 741.
\end{thebibliography}
\end{document}